# Automated detection of COVID-19 cases from chest X-ray images using deep neural network and XGBoost

Hamid Nasiri,[1] Sharif Hasani,[2]

[1] Department of Computer Engineering, Amirkabir University of Technology, Tehran, Iran
[2] Electrical and Computer Engineering Department, Semnan University, Semnan, Iran

Correspondence should be addressed to Hamid Nasiri; h.nasiri@aut.ac.ir

## Abstract

In late 2019 and after COVID-19 pandemic in the world, many researchers and scholars have tried to provide methods for detection of COVID-19 cases. Accordingly, this study focused on identifying COVID-19 cases from chest X-ray images. In this paper, a novel approach to diagnosing coronavirus disease from X-ray images was proposed. In the proposed method, DenseNet169 deep neural network was used to extract the features of X-ray images taken from the patients' chests and the extracted features were then given as input to the Extreme Gradient Boosting (XGBoost) algorithm so that it could perform the classification task. Evaluation of the proposed approach and its comparison with the methods presented in recent years revealed that the proposed method was more accurate and faster than the existing ones and had an acceptable performance in detection of COVID-19 cases from X-ray images.

## 1. Introduction

COVID-19 virus was reported in Wuhan, China, in late December 2019 with unknown causes, after which it spread rapidly throughout the world [1]–[3]. The virus spread in most parts of China over 30 days [4]. The infectious disease caused by this type of virus was named COVID-19 by the World Health Organization (WHO) on February 11, 2020 [2]. COVID-19 was reported in Iran on February 21, 2020. About 3.5 and 192 million confirmed cases were identified in Iran and throughout the world until July 20, 2021, respectively. Most types of coronavirus affect animals, but they can also be transmitted to humans because of their common natures. Severe Acute Respiratory Syndrome (SARS-CoV)-associated coronavirus causes a severe respiratory disease and death in humans [5]. The well-known signs and symptoms of COVID-19 include fever, cough, sore throat, headache, fatigue, muscle pain, and shortness of breath [6]. During this pandemic, COVID-19 virus has had direct impacts on the lifestyles of most communities, including humans' health, social welfare, businesses, and social relationships. It has also left indirect effects, such as reducing the quality of education in schools and universities, weakening family relationships, decreasing sports activities, and so on.

The most common method of COVID-19 diagnosis in individuals is the Real-Time Reverse Transcription-Polymerase Chain Reaction (RT-PCR) assay. However, identification with this approach is time-consuming and the results have a high level of false-negative errors [7], [8]. Alternatively, chest radiographic imaging methods, such as Computed Tomography scan (CT-scan) and X-ray, have a vital and effective role in the timely diagnosis and treatment of this disease [7], especially in pregnant women and children [9], [10]. Chest X-ray Radiograph (CXR) images are mostly utilized to diagnose chest pathology and are rarely applied to detect COVID-19.

This research was conducted on these types of images due to the presence of radiographic imaging devices in most hospitals and specialized clinics [9], [11]. According to previous studies, radiological images of patients with COVID-19 bear important and useful information for identifying the virus in the body [12]. One of the disadvantages of using CXR images is that they can detect soft tissues with a poor contrast and thus are not capable of determining the extent of a patient's lung involvement [13], [14]. To compensate for this defect, Computer-Aided Diagnosis (CAD) systems can be employed [15], [16]. Most CAD systems depend on the development of Graphic Processing Units (GPUs). GPUs in these systems are applied to implement medical image processing algorithms, such as image enhancement and limb or tumor segmentation [17], [18]. The development and spread of the use of artificial intelligence, especially some of its branches, such as machine learning and deep learning, have highly contributed to the greater intelligence of processes compared to humans' intelligence. Artificial intelligence has also had a significant impact on the speed of these processes in such fields as medical sciences for diagnosis or even treatment. For instance, in areas like lung [19], [20] and cardiovascular [21], [22] diseases and brain surgery [23], [24] it has contributed so effectively to the medical community and patients.

Advances in deep learning have shown good results on the ground of medical image analysis and radiology [25], [26]. Deep learning has miscellaneous architectures, each of which involves a variety of applications in their related fields. One type of deep learning architecture is Deep Convolutional Neural Network (DCNN), which is employed specifically in the field of image processing. Among its varied applications, pattern recognition and image classification could be mentioned [27].

Depending on the problem involved, DCNNs can be used in many different ways. One of the existing methods is the use of pre-trained neural networks, which were utilized in this research. Based on this approach, pre-trained models that are freely available are employed and image features are extracted using the deep neural networks. The second step after extracting image features is to use classification methods to conduct classification and prediction processes. Among the various classification approaches, such as Support Vector Machines (SVM), Decision Tree, etc., XGBoost classifier was applied in this paper. The rest of the paper is organized as follows: Section 2 describes related works. Section 3 explains necessary background concepts like XGBoost algorithm. In Section 4, the proposed method will be presented. In section 5, the experimental results are reported and analyzed. Finally, the conclusion will be presented in Section 6.

## 2. Related Works

Apostolopoulos et al. [28] carried out a study on a set of X-ray images from patients with pneumonia, COVID-19, and healthy individuals with the aim of assessing the Convolutional neural network (CNN) performance. In this research, transfer learning was utilized and the research process was done in 3 stages. The results demonstrated that the use of deep learning could lead to the extraction of significant features from COVID-19. Kaur et al. [29] proposed a modified AlexNet for feature extraction and classification. They employed Strength Pareto evolutionary algorithm-II for hyper parameters tunning and test their proposed model on a four-class dataset. In another study, Wang et al. [30] presented the COVID-Net network (a DCNN for COVID-19 detection), which was implemented on X-ray images. Their proposed network could help physicians during the screening phase. Alruwaili et al. [31] proposed an enhanced Inception-ResNetV2 deep learning model for detection of COVID-19 cases from chest X-ray images. They also used Grad-CAM algorithm to specify the infected regions of the lungs. In their research, Sethy et al. [32] used deep learning and SVM to detect coronavirus-involved patients using X-ray images. Since SVM is a powerful, it was applied in their classification process.

Fan et al. [33] employed transfer learning techniques for the identification of COVID-19 cases. They used five different deep neural networks (AlexNet, MobileNetv2, ShuffleNet, SqueezeNet, and Xception) with three different optimizers. They suggested MobileNetv2 with Adam optimizer as their best model. Hemdan et al. [34] proposed COVIDX-Net, which consists of networks, such as VGG16 and Google MobileNet. Mishra et al. [35] proposed a decision fusion based approach, which combines predictions from different DCNNs to identify COVID-19 from chest CT images. In their study, Narin et al. [36] proposed five models for diagnosing patients with pneumonia and coronavirus via X-ray images. These models were based on pre-trained CNNs, such as ResNet152, ResNet101, ResNet50, Inception-ResNetV2, and InceptionV3. Shorfuzzaman et al. [37] proposed a novel CNN based framework, which combines weights from different models to extract features and uses a custom classifier for classification. Sung et al. [38] developed a system to identify patients with COVID-19 by using the CT images collected from hospitals in two provinces of China. Li et al. [39] combined the Stack Generalization ensemble learning with the VGG16 deep neural network for classification of chest CT images.

In a study, Ozturk et al. [12] proposed a new model for detecting COVID-19 by using X-ray images. Their proposed model was presented based on the two problems of binary classification (to distinguish COVID-19 from no finding class) and multi-class classification (to distinguish COVID-19, pneumonia, and no finding classes). They proposed a DCNN called DarkNet, which included 17 convolutional layers. They achieved accuracies of 98.08 and 87.02% in their binary and multi-class classifications, respectively. For more details on other proposed methods that use deep learning for detection and diagnosis of COVID-19, please refer to [40].

## 3. Backgrounds

In this section, the XGBoost algorithm employed for classification in our proposed method will be introduced. XGBoost is an efficient and scalable algorithm based on tree boosting that was proposed by Chen & Guestrin in 2016 [41]. In fact, it is an improved version of the Gradient Boosted Decision Tree (GBDT) method that has been proven that does not have its computational limitations [42], [43]. Nevertheless, it differs from the GBDT method somehow. GBDT uses the first-order Taylor expansion, while the second-order Taylor expansion is utilized in the XGBoost's loss function. In addition, the objective function is normalized in XGBoost to alleviate the model's complexity and prevent it from overfitting [44].

Mathematically speaking, the output ($\hat{y}_i$) predicted by the XGBoost model is the sum of all the scores predicted by *K* Trees as shown in Equation1:

$$\hat{y}_i = \sum_{k=1}^{K} f_k(\mathbf{x}_i),\ f_k \in \Gamma \tag{1}$$

where $\mathbf{x}_i$ is the input feature vector; $f_k$ indicates the score of the $K$ th tree, which is also known as leaf score; $K$ denotes the number of regression trees; and $\Gamma$ stands for the function space that includes all possible regression trees [45], [46]. XGBoost tries to minimize the regularized objective function given in Equation 2 so as to learn the set of functions used in the model.

$$\phi = \sum_{i=1}^{n} l(y_i, \hat{y}_i) + \sum_{j=1}^{k} \Omega(f_j) \tag{2}$$

where $l(\hat{y}_i, y_i)$ measures the difference between the target value, $y_i$, and the predicted value, $\hat{y}_i$, and $\Omega(f)$ is the regularizing term that penalizes complex models to prevent overfitting. $\Omega(f)$ is calculated by Equation 3:

$$\Omega(f) = \gamma T + \frac{1}{2}\lambda \|\omega\| \tag{3}$$

where $T$ is the number of leaf nodes; $\omega$ is the score of each leaf; and $\gamma$, which is the cost of complexity, and $\lambda$, which is a hyper parameter that determines the model's degree of regularization, represent constant coefficients [46]–[48].

## 4. Proposed Method

Considering the past similar research activities, as well as the common methods of using artificial intelligence techniques in image processing, especially for medical images, our proposed method was to use the extracted features of images using pre-trained networks. One of the applications of artificial intelligence is the use of transfer learning techniques. In this technique, various networks are designed and trained with a huge set of available data and the weights of network layers are calculated. For example, in image processing, the ImageNet dataset contains millions of images in 1000 different classes. There are several methods to utilize pre-trained networks, which include:

    1- Using the structures of pre-designed networks to train one's own model, remove the last layer of the presented network, and finally add layers to perform classification.

    2- Extracting image features by using the pre-trained models and using the extracted features to perform classification via other algorithms.

In this paper, the features were extracted by using the second method and the XGBoost classifier was employed for classification. In this way, the images were first given as input to the DenseNet169 deep neural network so that the network could extract the image features. The extracted features were then given as input to the XGBoost algorithm to perform the classification operation. The framework of the proposed method can be seen in Figure 1.

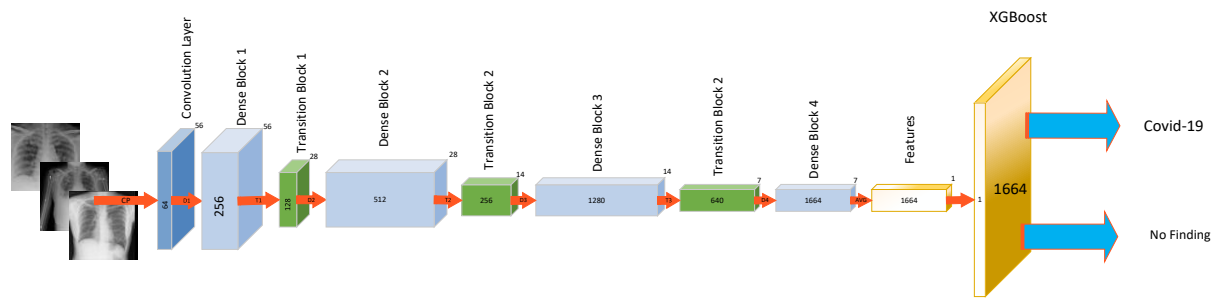

Figure 1: Framework of the proposed method

## 5. Experimental Results

The ChestX-ray8 dataset collected by Wang et al. [49] was used to evaluate the proposed method. This dataset contains 1125 X-ray images of the studied individuals' chests, including 125 images labeled as COVID-19, 500 images labeled as pneumonia, and 500 images labeled as no findings. Figure 2 depicts sample images in the dataset.

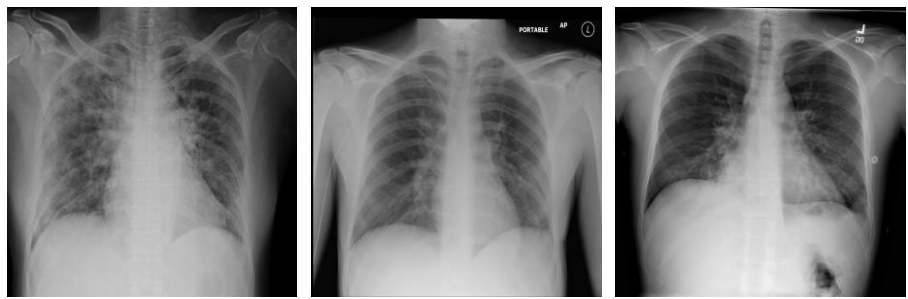

Figure 2: Samples of images in the dataset

This research was done in two phases. In the first phase, the best (pre-trained) DNN was selected to extract the features and in the second phase, the XGBoost classifier parameters are set by a trial-and-error. Also, the ChestX-ray8 dataset was used in two cases. Two-class problem, including COVID-19 and no findings (625 images) and three-class problem, including COVID-19, pneumonia, and no findings classes (1125 images).

In phase 1, 17 pre-trained neural networks were assessed and the XGBoost classifier was utilized along with the default parameters for classification. Table 1 shows the average accuracy of the DNNs for each of the two-class and three-class problems. It should be noted that a 5-fold cross-validation method was applied to obtain the average accuracy in this experiment.

Table 1 The average accuracy comparison of different DNNs

| DNN | Average Accuracy (%) | |
|---|---|---|
| | Three-class Problem | Two-class Problem |
| Xception | 78.84 | 93.59 |
| VGG16 | 81.68 | 96.48 |
| VGG19 | 80.08 | 95.36 |
| ResNet50 | 80.71 | 95.51 |
| ResNet152 | 79.55 | 95.68 |
| ResNet50V2 | 80.53 | 94.71 |
| ResNet101V2 | 76.88 | 93.95 |
| ResNet152V2 | 77.60 | 93.59 |
| InceptionV3 | 79.02 | 92.79 |
| InceptionResNetV2 | 68.44 | 90.72 |
| MobileNet | 79.55 | 95.51 |
| MobileNetV2 | 82.57 | 96.16 |
| DenseNet121 | 82.51 | 96.32 |
| DenseNet169 | **83.02** | **97.43** |
| DenseNet201 | 82.31 | 96.63 |
| NASNetMobile | 74.57 | 93.11 |
| EfficientNetB0 | 80.00 | 97.28 |

As can be seen, the DenseNet169 network had the best accuracy in both cases. As a result, this network was selected to extract the features of the proposed model in phase 2. The input to this network included images with the dimensions of 224×224×3 and its output consisted of 1664 features, which the network extracted from the given images. After determining the network type, the XGBoost classifier parameters were set. Table 2 shows the parameters used in the XGBoost algorithm.

Table 2 The XGBoost parameter settings

| Parameter | Value |
|---|---|
| Base Learner | Gradient boosted tree |
| Tree construction algorithm | Exact greedy |
| Number of gradients boosted trees | 100 |
| Learning rate ($\eta$) | 0.44 |
| Lagrange multiplier ($\gamma$) | 0 |
| Maximum depth of trees | 6 |

For two-class problem, 5-fold cross validation was used and for three-class problem, 80% of dataset was used for training and the remaining 20% was used as test set. The average accuracy for two-class problem was 98.23% and the test accuracy for three-class problem was 89.70%. The confusion matrices for each of the 5 folds in two-class problem are shown in Figure 3 and the confusion matrix for three-class problem is shown in Figure 4.

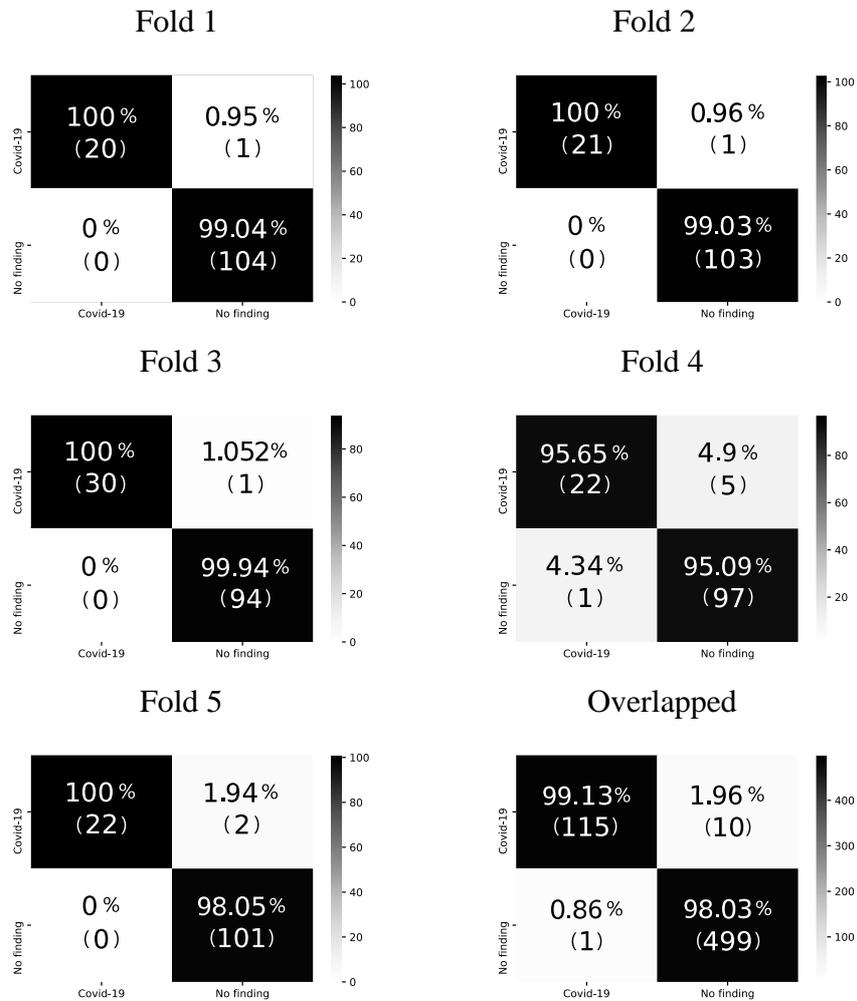

Figure 3: Confusion matrices for two-class problem

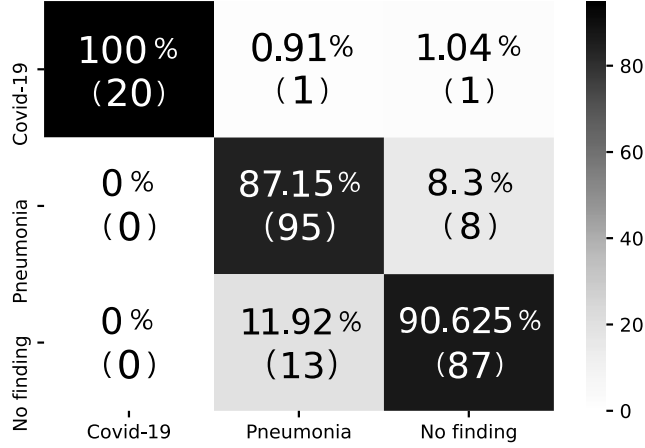

Figure 4: Confusion matrix for three-class problem

The results of comparing the proposed approach with the method proposed by Ozturk et al. [11] for the three-class and two-class problems can be seen in Tables 3 and 4, respectively. Also, a comparison of the results with other proposed methods is given in Table 5.

Table 3 Comparison of the proposed method with DarkCovidNet (Three-class problem)

|             | Proposed Method | DarkCovidNet |
|-------------|-----------------|--------------|
| Sensitivity | **95.20**       | 85.35        |
| Specificity | **100**         | 92.18        |
| Precision   | **92.50**       | 89.96        |
| F1-score    | **91.20**       | 87.37        |
| Accuracy    | **89.70**       | 87.02        |

Table 4 Comparison of the proposed method with DarkCovidNet (Two-class problem)

| Performance Metrics | | Fold 1 | Fold 2 | Fold 3 | Fold 4 | Fold 5 | Average |
|---|---|---|---|---|---|---|---|
| Sensitivity | Proposed Method | 95.20 | 95.40 | **96.70** | 81.40 | 91.40 | 92.08 |
|             | DarkCovidNet    | **100** | **96.42** | 90.47 | **93.75** | **93.18** | **95.13** |
| Specificity | Proposed Method | **100** | **100** | **100** | 89.90 | **100** | **99.78** |
|             | DarkCovidNet    | 100 | 96.42 | 90.47 | **93.75** | 93.18 | 95.30 |
| Precision   | Proposed Method | 99.50 | **99.50** | **99.40** | 95.30 | **99.02** | **98.54** |
|             | DarkCovidNet    | **100** | 94.52 | 98.14 | **98.57** | 98.58 | 98.03 |
| F1-score    | Proposed Method | 98.50 | **98.50** | **98.20** | 92.50 | **97.30** | **97.00** |
|             | DarkCovidNet    | **100** | 95.52 | 93.79 | **95.93** | 95.42 | 96.51 |
| Accuracy    | Proposed Method | 99.2 | **99.2** | **99.2** | 95.2 | **98.4** | **98.24** |
|             | DarkCovidNet    | **100** | 97.60 | 96.80 | **97.60** | 97.60 | 98.08 |

Table 5 Comparison of the proposed method with other deep learning-based methods

| Study | Type of Images | Number of Samples | Method Used | Accuracy (%) |
|---|---|---|---|---|
| Apostolopoulos et al. [28] | Chest X-ray | 1428 | VGG-19 | 93.48 |
| Wang et al. [30] | Chest X-ray | 13645 | COVID-Net | 92.40 |
| Sethy et al. [32] | Chest X-ray | 50 | ResNet50 + SVM | 95.38 |
| Hemdan et al. [34] | Chest X-ray | 50 | COVIDX-Net | 90.00 |
| Narin et al. [36] | Chest X-ray | 100 | Deep CNN ResNet-50 | 98.00 |
| Song et al. [38] | Chest CT | 1485 | DRE-Net | 86.00 |
| Wang et al. [50] | Chest CT | 453 | M-Inception | 82.90 |
| Zheng et al. [51] | Chest CT | 542 | UNet + 3D Deep Network | 90.80 |
| Xu et al. [52] | Chest CT | 443 | ResNet + Location Attention | 86.60 |
| Ozturk et al. [12] | Chest X-ray | 625 | DarkCovidNet | 98.08 |
|  |  | 1125 |  | 87.02 |
| Proposed Method | Chest X-ray | 625 | DenseNet169 + XGBoost | **98.23** |
|  |  | 1125 |  | **89.70** |

As can be seen in Tables 3 and 4, the proposed method had better performance than the DarkCovidNet network in both three-class and two-class problems. Noteworthy, the proposed approach had a higher speed and lower computational complexity than the method presented by Ozturk et al. due to the fact that it did not require training of the DNN. Note that the proposed method just trains the XGBoost algorithm. Table 5 demonstrates that the proposed method is more accurate than other deep learning-based models. However, it should be noted that the results presented in Table 5 were obtained from different datasets.

## 6. Conclusion

This study aimed to detect and identify people with COVID-19 with a focus on the use of non-clinical approaches and artificial intelligence techniques. In the proposed method, DenseNet169 was employed to extract image features and the XGBoost algorithm was used for classification. The obtained results revealed that the detection accuracy of the proposed method in the two-class problem was 98.24%, which was higher than other proposed methods. Also, 89.70% accuracy was reached in the three-class problem, thus indicating better performance compared to the DarkCovidNet network. Besides being highly accurate, the proposed approach had a higher speed and lower computational complexity than the other proposed methods due to the fact that it did not require training of DNN.

## Data Availability

Publicly available ChestX-ray8 dataset was used in this study, which is available at https://github.com/muhammedtalo/COVID-19/

## Code Availability

The source code of the proposed model required to reproduce the predictions and results is available at https://github.com/sharifhasani/CovidDetection

## Conflicts of Interest

The authors declare that they have no conflicts of interest.